\documentclass[journal]{IEEEtran} 

\usepackage{cite}
\usepackage{amsmath,amssymb,amsfonts}
\usepackage{algorithmic}
\usepackage{graphicx,color}
\usepackage{url}
\usepackage{booktabs}
\usepackage{textcomp}
\usepackage{multirow}
\usepackage[nolist]{acronym}
\usepackage[T1]{fontenc}

\def\BibTeX{{\rm B\kern-.05em{\sc i\kern-.025em b}\kern-.08em
    T\kern-.1667em\lower.7ex\hbox{E}\kern-.125emX}}

\begin{acronym}[]
    \acro{ML}{Machine Learning}
    \acro{CCI}{Co-channel Interference}
    \acro{SINR}{Signal-to-Interference-plus-Noise Ratio}
    \acro{V2X}{Vehicle-to-Anything communication}
    \acro{IoT}{Internet of Things}
    \acro{MIMO}{Multiple Input Multiple Output}
    \acro{CNN}{Convolutional Neural Network}
    \acro{RNN}{Recurrent Neural Networks}
    \acro{FLOPS}{Floating Point Operations Per Second}
    \acro{MACs}{Multiply-Accumulate Operations}
    \acro{RFIC}{Radio-Frequency Integrated Circuit}
    \acro{WSN}{Wireless Sensor Network}
    \acro{LSTM}{Long short-term memory}
    \acro{SOI}{Signal of Interest}
    \acro{QPSK}{Quadrature Phase Shift Keying}
    \acro{RRC}{Root Raised Cosine}
    \acro{QAT}{Quantization-Aware Training}
    \acro{MSE}{Mean Squared Error}
    \acro{BER}{Bit Error Rate}
    \acro{OOM}{Out-of-Memory}
\end{acronym}

\makeatletter
\def\ps@IEEEtitlepagestyle{%
  \def\@oddhead{%
    \parbox{\textwidth}{%
      \centering \footnotesize \copyright2024 IEEE. This work has been submitted to the IEEE for possible publication. \\
      Copyright may be transferred without notice, after which this version may no longer be accessible.
    }\vspace{1em}%
  }%
  \def\@oddfoot{}%
}
\makeatother

\begin{document}

\title{High-Throughput Blind Co-Channel Interference Cancellation for Edge Devices Using Depthwise Separable Convolutions, Quantization, and Pruning}

\author{
Mostafa~Naseri$^{\dagger}$ \quad  
Eli~De~Poorter$^{\dagger}$ \quad  
Ingrid~Moerman$^{\dagger}$ \quad 
H.~Vincent~Poor$^{\ddagger}$ \quad 
Adnan~Shahid$^{\dagger}$ \\ [0.1in]
$^{\dagger}$ IDLab, Department of Information Technology at Ghent University - imec \\
$^{\ddagger}$ Department of Electrical and Computer Engineering, Princeton University, Princeton
}

\maketitle



\begin{abstract}
Co-channel interference cancellation (CCI) is the process used to reduce interference from other signals using the same frequency channel, thereby enhancing the performance of wireless communication systems. An improvement to this approach is blind CCI, which reduces interference without relying on prior knowledge of the interfering signal characteristics. Recent work suggested using machine learning (ML) models for this purpose, but high-throughput ML solutions are still lacking, especially for edge devices with limited resources. This work explores the adaptation of U-Net Convolutional Neural Network models for high-throughput blind source separation. Our approach is established on architectural modifications, notably through quantization and the incorporation of depthwise separable convolution, to achieve a balance between computational efficiency and performance. Our results demonstrate that the proposed models achieve superior MSE scores when removing unknown interference sources from the signals while maintaining significantly lower computational complexity compared to baseline models. One of our proposed models is deeper and fully convolutional, while the other is shallower with a convolutional structure incorporating an LSTM. Depthwise separable convolution and quantization further reduce the memory footprint and computational demands, albeit with some performance trade-offs. Specifically, applying depthwise separable convolutions to the model with the LSTM results in only a 0.72\% degradation in MSE score while reducing MACs by 58.66\%. For the fully convolutional model, we observe a 0.63\% improvement in MSE score with even 61.10\% fewer MACs. Additionally, the models exhibit excellent scalability on GPUs, with the fully convolutional model achieving the highest symbol rates (up to 800$\times10^3$ symbol per second) at larger batch sizes. Overall, our findings underscore the feasibility of using optimized machine-learning models for interference cancellation in devices with limited resources. 
\end{abstract}


\maketitle

\section{Introduction}
\IEEEPARstart{S}{ignal} separation, an integral technique in signal processing, has a broad spectrum of applications across diverse fields. In audio signal processing, it plays a critical role in noise reduction for speech and music recordings, separating vocals from instrumentals, and enhancing clarity in hearing aids \cite{defossez2019music}. This technology not only improves the quality of audio content but also aids in accessible communication. In the realm of medical imaging, signal separation techniques are pivotal for artifact removal in MRI and CT scans, contributing to clearer and more accurate diagnostic images \cite{ASADZADEH2020108740}. It is also instrumental in analyzing complex signals in electroencephalography (EEG) and magnetoencephalography (MEG), enhancing our understanding of neural activities. Financial time series analysis also benefits from signal separation. It aids in filtering out market noise, allowing analysts to identify informative trends and underlying factors that influence asset prices, thus facilitating more informed investment decisions \cite{szupiluk2016analysis}. 
In the field of astronomy and space exploration, signal separation is key in filtering cosmic noise in radio telescope data and is integral in the signal processing for satellite communication and operations in the deep space network \cite{boulais2021unmixing}.
Furthermore, in video processing, signal separation techniques are used for noise removal and feature enhancement in surveillance footage and for separating and reconstructing overlaid images or videos, providing clearer visual data for various applications.

In the evolving landscape of wireless communications, \ac{CCI} emerges as an interesting challenge, especially due to the increasing demand for spectral efficiency and the need for high-quality signal transmission. Primarily caused by spectral reuse—a practice where the same frequency bands are deployed across different transmitters—\ac{CCI} significantly compromises the \ac{SINR}, a crucial determinant of communication reliability and quality.

As wireless technologies spread across various platforms, from satellites to terrestrial vehicles, and as edge devices become widespread, the issue of \ac{CCI} grows more severe.
These devices frequently encounter a stream of composite signals, composed of both desired and undesired transmissions, which severely degrade the \ac{SINR}. Effective interference cancellation mechanisms are therefore essential to process and extract the intended signals from this interference, enhancing signal clarity and overall communication fidelity. Furthermore, the concept of underlay communication, particularly relevant in the context of next-generation 5G networks, highlights another critical application of interference cancellation~\cite{9801546, 10356263, 9529215}. Underlay strategies allow secondary users—such as unmanned aerial systems and remote sensors—to operate within the same spectral and temporal resources as primary users. This method not only facilitates innovative uses of existing spectral resources but also significantly boosts spectrum efficiency, showcasing the profound relevance of advanced \ac{CCI} cancellation techniques in modern telecommunication systems.

The importance of addressing \ac{CCI} in modern communication systems is driven by several key technological advancements and emerging market demands. Notably, the growth of safety-critical systems \cite{silva2023comprehensive}, such as those used in automotive and healthcare applications, necessitates highly reliable communications even in environments disturbed by strong interference. This has encouraged interest in technologies like successive interference cancellation, which can dramatically enhance the quality and reliability of signal reception in congested scenarios \cite{9529215}. Furthermore, the expansion of \ac{IoT} devices, which often operate in dense networks with limited spectral resources, calls for more sophisticated interference management techniques. \ac{V2X} communications, crucial for the safety of autonomous and connected vehicles, also underscore the need for robust interference cancellation mechanisms to ensure uninterrupted communication in dynamic environments.

In addition, there is growing interest in exploiting unlicensed spectrum and cognitive radio networks to enhance spectrum efficiency. These technologies rely heavily on advanced interference cancellation to prevent disruption of existing services while supporting new wireless applications. Satellite communications, another area experiencing rapid growth due to global connectivity initiatives, face unique challenges related to \ac{CCI} due to the overlapping signals from multiple sources.
Together, these factors make the study of \ac{CCI} cancellation techniques not only relevant but essential for the next generation of wireless systems, driving forward innovations that will shape the future of global communications.
As the complexity and density of wireless networks continue to grow, traditional signal processing approaches struggle to keep up with the dynamic and unpredictable nature of modern interference patterns. These methods often rely on predefined models and assumptions that may not accurately capture real-world conditions, leading to suboptimal performance in practical scenarios.
While the integration of machine learning models into interference cancellation strategies marks a progressive step toward smarter communication systems \cite{naseri2024u}, it also unveils significant challenges. Specifically, the computational limitations of nodes within modern networks pose a considerable bottleneck. Most existing machine learning algorithms for \ac{CCI} cancellation demand high computational resources, which can be impractical in real-time applications where quick processing is critical \cite{henneke2024improving, yapar2024demucs}. This paper addresses these crucial gaps by proposing and investigating efficient neural network architectures and techniques that are tailored for real-time interference cancellation without compromising performance. This approach seeks to bridge the gap between the theoretical potential of machine learning models and their practical implementation in communication networks disurbed by \ac{CCI}.
\subsection{Background on Blind Co-Channel Interference}
The history of \ac{CCI} management is marked by continuous evolution, beginning with simple linear filters in analog signal processing. These filters, albeit basic, laid the groundwork for more sophisticated non-linear and predictive methods \cite{6596755}. With digital signal processing, adaptive filters emerged, dynamically adjusting to signal environments and outperforming their analog predecessors. The spread of spectrum technology, including frequency-hopping and CDMA, offered robust solutions to reduce \ac{CCI} by dispersing the signal across a broader bandwidth. The implementation of \ac{MIMO} technology marked a significant leap. Utilizing multiple antennas at both transmission and reception points, MIMO systems harnessed spatial diversity and multiplexing to enhance capacity and mitigate interference \cite{CLERCKX2013419}.

The advent of machine learning, especially deep learning techniques like \acp{CNN} and \acp{RNN}, promises a new era in \ac{CCI} management. These networks excel in modeling complex interference patterns and separating sources in densely populated signal environments. However, most \ac{ML}-based \ac{CCI} studies focus on known interference types, as highlighted in \cite{jayashankar2024data}. Notably, models like WaveNet, though efficient in handling known interference \cite{jayashankar2024data}, exhibit significant performance degradation in blind or unknown interference scenarios \cite{naseri2024ffcunet}. 

Furthermore, in the realistic conditions of wireless networks, where we frequently encounter interference with unknown characteristics, the ability to cope effectively becomes essential. This necessity leads us to explore the realm of blind \ac{CCI} cancellation. In such a scenario, our methods must proficiently address interference without prior insight into its specific properties.
\subsection{Challenges in Real-Time Interference Cancellation}

Achieving real-time interference cancellation in wireless communications encompasses several key aspects: inference time, throughput, and consistency. Inference time, particularly, refers to the time taken by a model to process input and generate output, exclusive of delays caused by data acquisition, pre/post-processing, and other systemic components. In this study, we specifically focus on the inference time as a crucial factor in determining real-time capability.

The primary challenges in achieving real-time interference cancellation are multi-fold. First and foremost is the computational complexity of the models. Advanced algorithms capable of effective interference cancellation often require significant computational resources, which can lead to increased processing time. In addition to processing power, memory constraints in edge devices limit the size and complexity of deployable models, thus posing a challenge to maintaining high performance under limited resources. Another critical challenge is power consumption. Edge devices, designed for efficiency and prolonged operation, necessitate models that are not only fast but also energy-efficient. This becomes a delicate balance, as increased model complexity and processing speed often translate into higher power consumption.

Previous approaches to interference cancellation have predominantly focused on performance metrics such as accuracy, often neglecting the aspect of inference time. For instance, we will show that models like WaveNet, despite having moderate \ac{FLOPS}, exhibit notably low processing speeds, likely due to the implementation specifics of their dilated convolutions. This disparity highlights the necessity of considering inference speed.

The importance of real-time processing in \ac{CCI} cancellation cannot be overstated. In many practical scenarios, especially in wireless communication, the delay sensitivity of interference cancellation and bit decoding makes the feasibility of machine learning models dependent on their real-time performance capabilities. Delays can significantly undermine the practicality of deploying these models in real-world scenarios. By designing a high-speed implementation, we aim to reduce the inference delay, thereby making real-time interference cancellation feasible on edge devices. This directly tackles the challenges of computational complexity, memory constraints, and power efficiency, paving the way for practical and efficient machine learning-based \ac{CCI} cancellation in resource-limited environments.

\subsection{Objective of the Study}

The primary objective of this study is to advance the implementation of real-time, \ac{CNN}-based models for \ac{CCI} cancellation. This research aims to bridge the gap between high computational demand and real-time processing requirements, especially in scenarios involving edge devices and mobile communications. To achieve this, we thoroughly examine the relationship between inference time, computational complexity, and memory footprint of these models.

Specifically, our study involves a detailed exploration of both full-precision and quantized (qint8) models. The choice of quantization aligns with our goal of enabling efficient \ac{CCI} cancellation on a broader spectrum of devices, including those with limited computational resources.

In terms of performance metrics, we employ symbols per second decoded by the model as a measure of inference speed, \ac{MACs} for computational complexity, and the number of model parameters to assess memory footprint. These metrics collectively offer a comprehensive view of the model's efficiency and practicality in real-world applications.

The expected outcome of this research is to enable higher quality communication for mobile and edge devices through improved signal quality post-\ac{CCI} cancellation. This improvement has the potential to reduce power consumption, especially in systems with battery constraints, making it particularly relevant for \acp{WSN} as well as other mobile communications scenarios. Additionally, the possibility of integrating \ac{CCI} cancellation with decoding processes offers a promising avenue for enhancing \acp{RFIC}, contributing to more energy-efficient and robust communication systems.

Mainly, this study has broader implications for energy consumption in mobile and edge devices. By optimizing CNN models for \ac{CCI} cancellation, we can substantially lower the energy requirements of these devices, making a significant contribution to the sustainability and efficiency of wireless communication networks globally.
The main contributions of this paper are: 
\begin{itemize}
\item Design of a set of convolutional neural networks for \ac{CCI} cancellation that outperforms the baseline models from the scientific literature.
\item Improvement of the proposed model architectures through optimization techniques for better performance.
\item Development of a fully convolutional architecture that is independent of input size.
\item Comprehensive evaluation of the models on both CPU and GPU, testing the scalability on GPU.
\end{itemize}
The remainder of this paper is structured as follows. In Section \ref{sec:sys_model}, we present the system model and discuss the error metric and loss function used. Section \ref{sec:methodology} outlines our methodology, detailing the model architecture, including the U-Net CNN model, depthwise convolution, and baseline models, as well as the quantization techniques employed. We also describe the dataset, data preparation, training process, and evaluation metrics. Section \ref{sec:results} provides the results, highlighting model performance, complexity, inference rate, comparative analysis, GPU parallelizability, and the impact of pruning. In Section \ref{sec:discussion}, we analyze the results and discuss the implications of our findings, along with future work directions. Finally, Section \ref{sec:conclusion} concludes the paper.
\section{System Model}
\label{sec:sys_model}
In our study, we consider a signal model where the received signal $y = s + b \in \mathbb{C}^L$ is a complex-valued mixture of length $L$. This mixture comprises the \ac{SOI} $s$ and an interference signal $b$, where $s$ and $b$ overlap in both time and frequency. The \ac{SOI}, $s$, is modeled as a \ac{QPSK} signal shaped with a \ac{RRC} waveform. This setup is inspired by the configurations of the recent data-driven signal separation competition in radio spectrum: ICASSP 2024 grand challenge \cite{jayashankar2024data}.
\subsection{Signal Model}
The $n$-th sample of $s$ is expressed as:
\begin{equation}
    s[n] = \sum_{k=0}^{L-1} a_k g[n - kF - \tau_0],
\end{equation}
where $F \in \mathbb{N}$ represents the symbol interval in discrete-time, and $\tau_0 \in \{0, \ldots, F-1\}$ denotes the offset applied to the first symbol. For our configuration, we set $F = 16$ and $\tau_0 = 8$. The function $g[n]$ is the discrete-time impulse response of the transmitter filter, which is a \ac{RRC} filter with a roll-off factor of 0.5 and a window length of 127 samples, i.e. 8 symbols.

\begin{figure}[h]
    \centering
    \includegraphics[width=1\columnwidth]{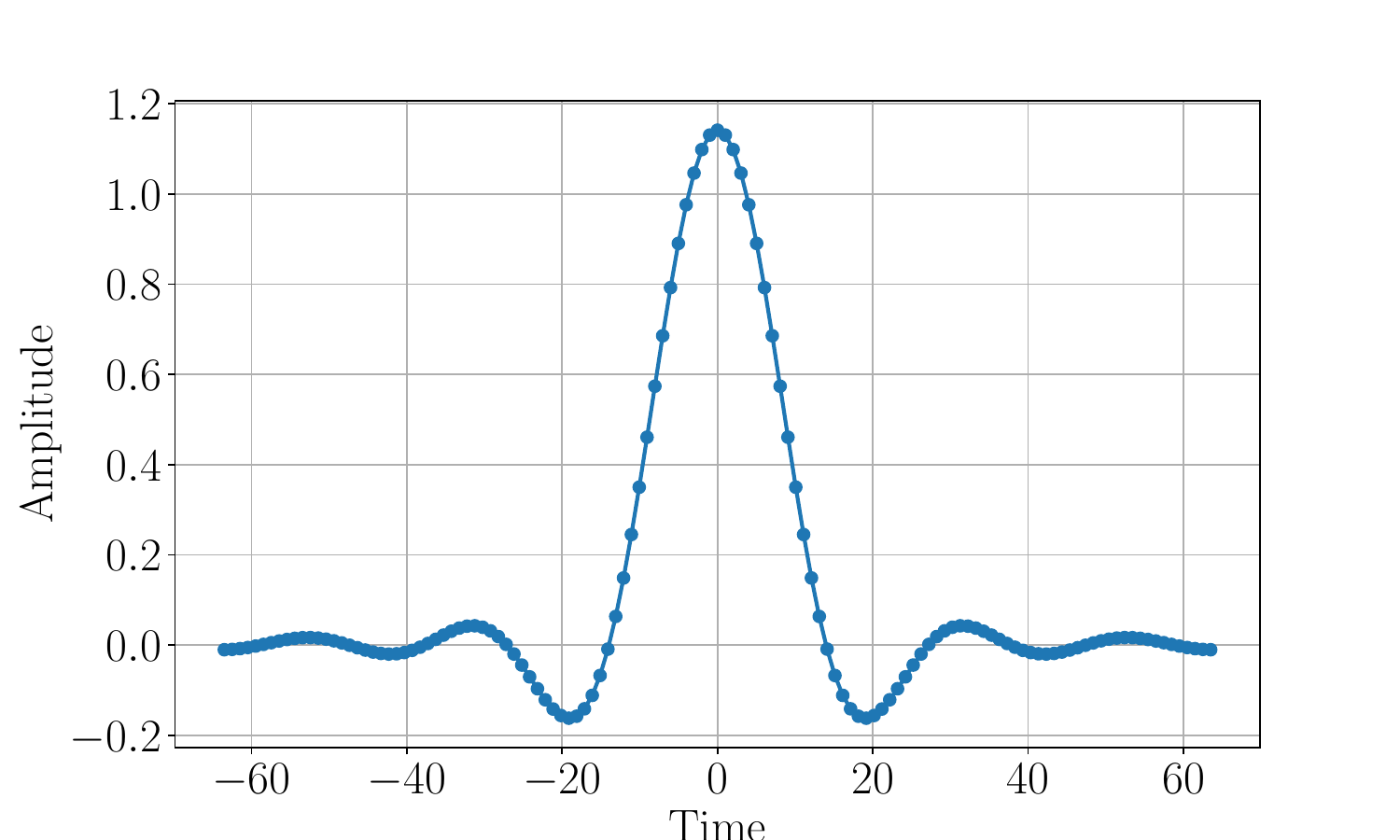}
    \caption{Root-Raised-Cosine Pulse Shaping Function}
    \label{fig:root_raised_cosine_pulse_shaping_function}
\end{figure}

In this model, $a_k$ denotes the QPSK symbols, which are mapped from the input bits. Initially, the bits are mapped to \ac{QPSK} symbols and subsequently passed through the pulse shaping filter to produce the signal $s[n]$. The interference signal $b$ originates from one of four datasets, representing different types of interference. The exact characteristics of the interference signals are generally unknown, as the task requires blind interference mitigation. The \ac{SINR} ranges from -30 dB to 0 dB in our evaluations. The received signal $y$ is modeled as a combination of the QPSK signal of interest and the interference:
\[
y[n] = s[n] + b[n], \quad n = 0, 1, \ldots, L-1.
\]

The objective is to extract the signal of interest $s$ from the received mixture $y$, despite the presence of unknown interference $b$.
This interference signal $b$ originates from one of four datasets used in the competition, and as such, the exact characteristics of the interference were unknown even to us.
This signal model provides the foundation for the signal processing and detection algorithms developed to achieve effective separation and mitigation of interference, ensuring robust communication in challenging environments.

\subsection{Error Metric and Loss Function}
\label{sub:error_metrics}
To evaluate the performance of our system, we use the \ac{MSE} between the true signal $s[n]$ and its estimated version $\hat{s}[n]$. The \ac{MSE} is defined as:
\[
\text{MSE} = \frac{1}{L} \sum_{n=0}^{L-1} (s[n] - \hat{s}[n])^2.
\]

Additionally, we introduce an \ac{MSE} score based on the truncated negative logarithmic value of the \ac{MSE}. This truncation is applied because \ac{MSE} values better than a certain threshold do not significantly enhance the ultimate goal of interference cancellation, which is bit decoding. By capping the \ac{MSE} score, we prevent the model from focusing on overly precise sample reconstructions in some examples and poor reconstruction in other that may lead to high \ac{MSE} score and poor average \ac{BER}. Instead, the model is encouraged to achieve uniformly good reconstructions across all samples, improving the overall \ac{BER}. In our work, we set the threshold at -50 dB \cite{jayashankar2024data}. 

Although we use this \ac{MSE} score for validation and testing, during training, we employ a smoothed version of this loss function to facilitate stable and continuous training. The smoothing function used is a sigmoid-type function, which helps to avoid discontinuities in the loss landscape. This approach ensures that our model not only achieves low \ac{MSE} but also maintains a reconstruction quality sufficient to enhance \ac{BER}, thereby achieving effective interference mitigation and robust communication. In the sequel, we solely focus on the MSE score, and a \ac{BER} analysis that includes bit decoding is left open for future research and studies.

\section{Methodology}
\label{sec:methodology}
In this study, we adopt a structured approach to address the challenge of \ac{CCI} cancellation by proposing two neural network architectures. Our methodology centers around the utilization of the U-Net architecture, renowned for its effectiveness in various signal and image processing tasks \cite{8513676, cha2023dnoisenet}. We propose two distinct variants of the U-Net model tailored specifically for \ac{CCI} cancellation: a shallow U-Net characterized by small numbers of filters at each layer and an \ac{LSTM} at the bottleneck (M1), and a deeper U-Net equipped with a larger number of filters (M2).

To enhance the performance and efficiency of these models, we introduce two significant modifications: the integration of depthwise convolutions and the application of quantization techniques. Depthwise convolutions are employed to reduce the computational demand while maintaining model effectiveness, and quantization is applied to further enhance the operational efficiency, making the models more suitable for deployment in real-time communication systems.

Each modified model's performance is systematically compared against baseline models, which will be introduced and described in subsequent sections. This comparative analysis aims to highlight the advantages of our proposed solutions in terms of both efficiency and effectiveness in \ac{CCI} cancellation.

    \subsection{Model Architectures}
    The architecture of the proposed models is guided by a set of design principles aimed at optimizing performance for CCI cancellation while maintaining computational efficiency and robustness. The development process involved a meticulous hyperparameter search to achieve an optimal balance between model complexity and performance. Starting with highly complex models, the number of parameters was progressively reduced. This parameter tuning was carefully managed to prevent any significant drop in performance, ensuring that both variants of the U-Net model—referred to as M1 and M2—are lightweight yet effective.

    \textbf{Efficiency and Scalability:} Both models were designed to be scalable and adaptable to different data types and input sizes. M2, being fully convolutional, showcases exceptional flexibility, allowing it to handle various input lengths during inference and to be retrained with different datasets without structural modifications. This design choice provides substantial advantages in terms of deployment in diverse operational environments.
    In contrast, M1 can practically handle different input lengths, but unlike M2, significant deviations from the input size used in training can degrade its performance (due to stateful nature of \ac{LSTM}).
    
    \textbf{Robustness:} To enhance the models' robustness, data augmentation techniques were employed extensively. These techniques, which will be detailed in \ref{subsec:Dataset} section, help the model generalize better across different interference scenarios. Unlike the approach taken in \cite{jayashankar2024data}, where separate models were trained for each type of interference, a single model in this study is trained across all available data types, which enhances its applicability and efficiency in real-world settings.
    
    \textbf{Temporal and Spatial Feature Extraction:} The M1 model incorporates an \ac{LSTM} layer at its bottleneck, focusing on capturing temporal dependencies and long-range dynamics within the signal. This is particularly beneficial for scenarios where the interference exhibits time-dependent characteristics. In contrast, M2 prioritizes spatial feature extraction through a network structure that is deeper and has a larger number of filters. This approach is designed to exploit the intricate spatial patterns of signals, making it suitable for environments with complex interference structures.

Two critical modifications, depthwise convolutions and quantization, have been implemented to optimize the performance and efficiency of the proposed neural network architectures.

\textbf{Depthwise Convolutions:} The introduction of depthwise convolutions in the models marks a significant advancement in reducing computational complexity \cite{howard2017mobilenets}. By performing convolutions in a separable manner—where the spatial and depth (channel) convolutions are decoupled—the number of trainable parameters and the computational demand, measured in \ac{MACs}, are substantially lowered. Although the actual inference speed gains from depthwise convolutions can vary depending on the platform, this modification inherently decreases the computational load, making the models more adaptable to a variety of deployment scenarios, including those with limited processing capabilities.

\textbf{Quantization:} The application of quantization techniques aims to further reduce the model size and computational requirements. By lowering the precision of the numerical values used in the model, quantization not only shrinks the overall memory footprint but also enhances the feasibility of implementing these models on processors with lower precision capabilities. This adjustment is crucial for deploying advanced neural network architectures in real-time communication systems where hardware efficiency is paramount.

    \subsubsection{U-Net CNN Model}
    
    The first U-Net Model, referred to as M1, is structured to effectively address the challenges of \ac{CCI} cancellation through a series of encoding and decoding layers, supplemented by a specialized \ac{LSTM} layer at its bottleneck.
    
    \textbf{M1 Encoder:} The encoder part of M1 consists of three layers, each equipped with a double convolution mechanism to enhance feature extraction. The number of filters in these layers scales up through [64, 128, 256] with corresponding strides [1, 2, 2], allowing the model to capture features from the input signal at various resolutions. in all convolutional layers kernel size of 3 has been used. Each encoder layer follows this sequence of operations:
    \begin{verbatim}
    x = conv1(x)
    x = GN1(x)
    x = relu1(x)
    x = conv2(x)
    x = GN2(x)
    x = relu2(x)
    \end{verbatim}
    where the details will be explained in the sequel.
    
    \textbf{M1 Bottleneck:} At the core of the architecture, an \ac{LSTM} layer with a hidden size of 64 serves as the bottleneck. This layer is crucial for capturing and integrating temporal dynamics of the signals, which is particularly vital for scenarios where interference exhibits time-dependent behaviors.
    
\begin{figure}[h]
        \centering
        \includegraphics[width=1\columnwidth]{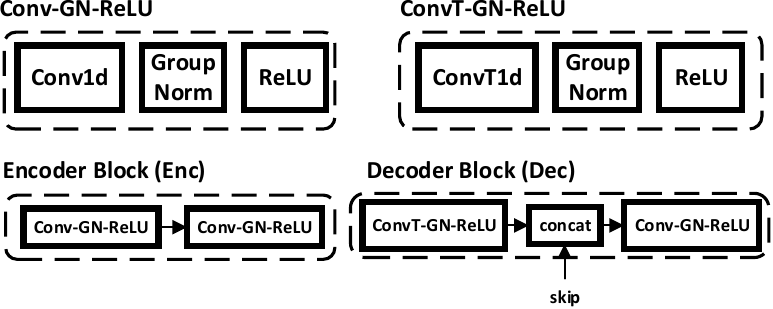}
        \caption{The architecture of the encoder and decoder blocks used in the models. The encoder block (Enc) consists of Conv1d, Group Normalization (Group Norm), and ReLU activation. The decoder block (Dec) mirrors the encoder but uses ConvT1d instead of Conv1d and includes concatenation (concat) for skip connections.}

        \label{fig:EncDec_Blocks}
    \end{figure}

    \textbf{M1 Decoder:} The decoder reverses the process of the encoder with two layers that progressively reconstruct the target signal from the obtained feature representations. The decoder layers include operations to combine the features from the encoder using skip connections and are structured as follows:
    \begin{verbatim}
    x = convT(x)
    x = GN(x)
    x = relu1(x)
    x = skip_handler(x, skip)
    x = conv(x)
    x = GN(x)
    x = relu2(x)
    \end{verbatim}
The encoder and decoder blocks, which form the fundamental building components of our models, are illustrated in Figure \ref{fig:EncDec_Blocks}.
    
    \begin{itemize}
    \item \textbf{Conv1d}: Applies a 1-dimensional convolution over an input signal composed of several input planes. This operation effectively captures local dependencies within the input data.
    \item \textbf{GroupNorm}: Applies Group Normalization over a mini-batch of inputs. This normalization method divides the channels into groups (here we use groups of 4) and computes within each group the mean and variance for normalization, which stabilizes the learning process and improves the training dynamics.
    \item \textbf{LeakyReLU}: Applies the Leaky Rectified Linear Unit function element-wise. Unlike the standard ReLU, LeakyReLU allows a small, non-zero gradient when the unit is not active, which helps prevent neurons from dying during training.
    \item \textbf{ConvTranspose1d}: Applies a 1-dimensional transposed convolution operator over an input image. Commonly referred to as deconvolution, it is instrumental in upsampling the feature maps to higher resolutions.
    \item \textbf{Skip Handler}: Manages the skip connections between the encoder and decoder blocks. Depending on the configuration (concatenation and addition are most common operations), we use concatenation in our models, which is crucial for integrating learned features at various levels and recovering spatial information lost during downsampling.
    \end{itemize}
        
        These operations collectively contribute to the robust feature extraction and reconstruction capabilities of the U-Net CNN Model, ensuring detailed and accurate signal processing critical for effective interference cancellation.
    
    \textbf{M2 Architecture:} The M2 model follows a similar architectural pattern to M1 but with key variations aimed at exploring different aspects of feature processing for \ac{CCI} cancellation. Specifically, M2 omits the \ac{LSTM} bottleneck and adjusts its layering to deepen the feature extraction capability.
    Both the encoder and decoder in M2 utilize the same double convolution structure as outlined for M1, but with modifications to the filter configuration and stride. The encoder features an extended sequence of layers with filters configured as [64, 128, 128, 128, 128, 128, 128, 128, 128] and corresponding strides of [1, 2, 2, 2, 2, 2, 2, 2, 2]. This configuration allows for deeper feature extraction while maintaining the model's efficiency through controlled spatial reduction in each strided convolution layer.
    
\begin{figure}[h]
        \centering
        \includegraphics[width=1\columnwidth]{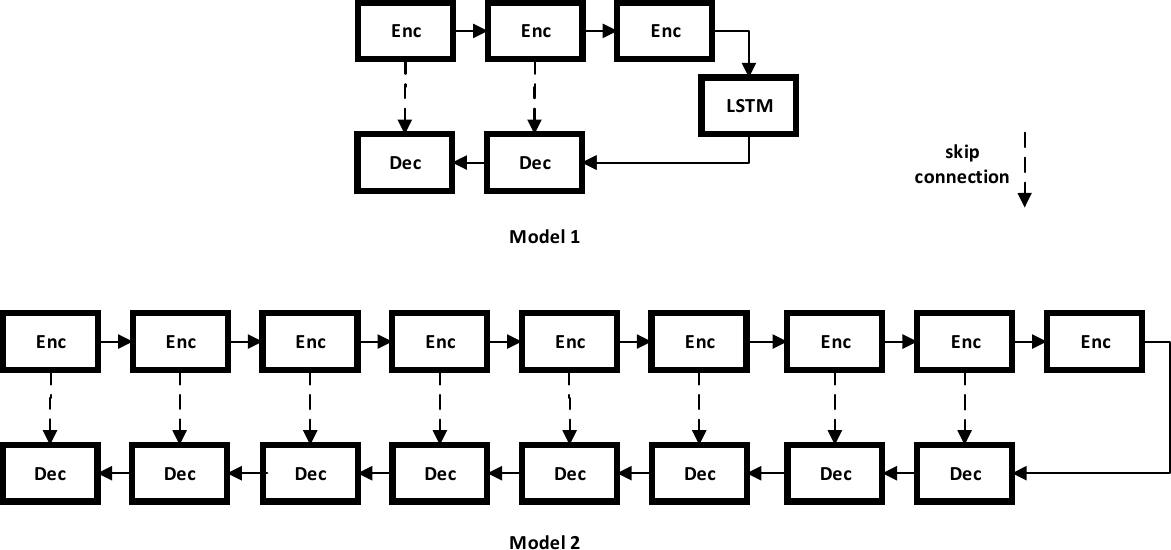}
        \caption{Model 1 (M1) utilizes a combination of encoder and decoder blocks with an LSTM layer and skip connections. Model 2 (M2) features a more extensive series of encoder and decoder blocks, incorporating multiple layers of skip connections for enhanced depth and complexity.}

        \label{fig:M1M2}
    \end{figure}

    In both M1 and M2 models, the first convolution in each double convolution block of the encoder and decoder is responsible for adjusting the number of channels and applying strided convolution to reduce spatial or temporal dimensions. The second convolution focuses on enhancing feature representation without changing the dimensions.
    By excluding the \ac{LSTM} layer, M2 emphasizes a purely convolutional approach.
    Figure \ref{fig:M1M2} presents the overall architecture of M1 and M2 models, showcasing the arrangement and integration of the encoder and decoder blocks with additional layers and skip connections.

        \subsubsection{Depthwise Convolution}
The depthwise convolution technique is integrated into the proposed models to optimize computational efficiency while maintaining the effectiveness of the feature extraction process. 
A depthwise convolution performs spatial convolution independently over each input channel, reducing the computational load compared to standard convolutions \cite{howard2017mobilenets}.
This section details the structure of depthwise convolution blocks in both encoder and decoder segments of the network.

\textbf{Encoder with Depthwise and Point-wise Convolutions:} The encoder utilizes depthwise convolution followed by point-wise convolution to efficiently process the input signal. This arrangement allows for a significant reduction in computational complexity by separating the filtering and combining aspects of the convolution process:
    \begin{verbatim}
    x = depthwise_conv(x)
    x = pointwise_conv(x)
    x = GN(x)
    x = relu1(x)
    x = depthwise_conv2(x)
    x = pointwise_conv2(x)
    x = GN(x)
    x = relu2(x)
    \end{verbatim}
The first depthwise convolution applies a spatial filter to each input channel independently, followed by a point-wise convolution that combines these filtered channels into the desired number of output channels. This two-step process efficiently manages feature extraction without excessively increasing the parameter count.

\textbf{Decoder with Depthwise and Point-wise Convolutions:} Similarly, the decoder employs depthwise and point-wise convolutions to reconstruct the signal from encoded features. It includes an additional step to handle the concatenated skip connections from the encoder, ensuring that spatial information is effectively reintegrated:
    \begin{verbatim}
    x = depthwise_convT(x)
    x = pointwise_convT(x)
    x = GN(x)
    x = relu1(x)
    x = skip_handler(x, skip)
    x = depthwise_conv(x)
    x = pointwise_conv(x)
    x = GN(x)
    x = relu2(x)
    \end{verbatim}
In the decoder, the depthwise convolutions first upsample the feature maps, followed by point-wise convolutions that adjust the channel dimensions to match the subsequent layers. This method effectively preserves detailed features while efficiently managing computational resources.
\subsubsection{Baseline Models from Prior Scientific Work}
To evaluate the effectiveness of the proposed U-Net architectures, comparisons are made against two well-established baseline models in the field of signal separation: WaveNet and ConvTasNet. These models were chosen due to their proven capabilities and distinct approaches to handling similar tasks in both audio and RF signal processing domains.

\textbf{WaveNet:} Originally developed for audio synthesis and later adapted for source separation, WaveNet is known for its utilization of dilated convolutions, which allow it to capture a wide range of temporal contexts efficiently. In our study, the WaveNet baseline is configured with 30 residual layers, using 128 residual channels, and a dilation cycle length of 10. This setup is designed to optimize the model's ability to separate overlapping signals by expanding its receptive field without significant increases in computational complexity. This model also forms the main baseline performance to beat in ICASSP 2024 grand challenge \cite{jayashankar2024data} for known interference types.

\textbf{ConvTasNet:} ConvTasNet employs a convolutional approach to perform source separation through a masking technique, distinguishing it from other spectral-domain separation methods. For our purposes, ConvTasNet is configured with parameters as follows: $N=512$, $L=16$, $B=128$, $H=512$, $P=3$, $X=8$, $R=3$. These parameters define the network’s architecture, including the number of filters ($N$), the length of the filters in samples ($L$), bottleneck width ($B$), number of channels in convolutional blocks ($H$), kernel size of the convolutional blocks ($P$), number of convolutional blocks within each repeat ($X$), and the number of repeats ($R$). Detailed explanations of these parameters can be found in \cite{luo2019conv} as well as in the Conv-TasNet implementation on GitHub\footnote{\url{https://github.com/JusperLee/Conv-TasNet}}.

    \subsection{Quantization Techniques}
    
    \ac{QAT} is employed to optimize the proposed models for efficient deployment, particularly focusing on convolutional layers. \ac{QAT} simulates the quantization effects during training, incorporating fake quantization nodes within the model architecture. This allows the training to adjust for the quantization-induced errors and refine the model parameters accordingly.
    
    \textbf{Implementation of \ac{QAT}:} In our methodology, quantization is applied only to the convolutional layers using an 8-bit representation. This approach is selected due to its balance between performance and computational efficiency. The \ac{QAT} process begins with the pre-trained unquantized model, which undergoes fine-tuning with quantization emulations activated. This strategy leverages the pre-existing model weights and optimizes them under quantization constraints, which avoids the need for training from scratch and ensures the model's performance remains robust.
    The primary benefits of applying \ac{QAT} include reduced model size and computational demands, making the models suitable for deployment on devices with limited computational resources (8 bit processors). Moreover, by incorporating quantization during the training phase, the models are better prepared for the reduced precision environment as compared to post-training quantization, which enhances their operational efficiency.
    
    \subsection{Dataset and Training}
    \label{subsec:Dataset}
    
    \subsubsection{Dataset Description}
    The datasets utilized in this study are sourced from the ICASSP 2024 grand challenge \cite{jayashankar2024data}, which offers a diverse set of interference types tailored for research in \ac{CCI} cancellation. The dataset comprises four distinct interference categories:
    \begin{itemize}
        \item \textbf{EMISignal1:} Electromagnetic interference from man-made sources.
        \item \textbf{CommSignal2} and \textbf{CommSignal3:} Digital communication signals from commercial wireless devices.
        \item \textbf{CommSignal5G1:} A 5G-compliant waveform.
    \end{itemize}
    The variety of these datasets provides a robust testing ground for the study, as it covers a broad spectrum of potential real-world interference scenarios. Notably, the provided labels indicating interference type are not utilized in our models to adhere to the study's assumption of unknown interference sources.
    
    \subsubsection{Data Preparation}
    Data preparation involves converting complex-valued 1D signals into two real-valued channels, representing the in-phase (I) and quadrature (Q) components. Additionally, two forms of data augmentation are employed to enhance the robustness of the models under varying signal conditions:
    \begin{enumerate}
        \item \textbf{Random SINR adjustment:} The \ac{SINR} is varied randomly between -30 dB and 0 dB by adjusting the gain of the interference signal.
        \item \textbf{Random Phase Shift:} A random phase between \(-\pi\) and \(\pi\) is added to the interference signal's phase before it is combined with the \ac{SOI}.
    \end{enumerate}
    The dataset distribution is detailed in terms of the number of superframes and the number of samples in each superframe for each type of interference, formatted as (number of superframes: number of samples):
    \begin{itemize}
        \item \textbf{CommSignal3:} 139: 260,000
        \item \textbf{CommSignal2:} 100: 43,560
        \item \textbf{CommSignal5G1:} 149: 230,000
        \item \textbf{EMISignal1:} 530: 230,000
    \end{itemize}
    
    A superframe is defined as the longest measurement window available in the dataset, from which smaller windows of length $L$ are extracted for training and validation. Additionally, 50 superframes are explicitly reserved and separated from the training dataset for the sole purpose of final testing, ensuring an unbiased evaluation of the model performance.
    For training purposes, the training dataset is randomly divided into an 80\% training set and a 20\% validation set, based on the number of superframes. Each batch for training or validation is dynamically generated by randomly selecting a superframe from the dataset, choosing a random starting index within the superframe, and extracting a segment of length \(L\).
     \subsubsection{Training Process}
    The training of the models is conducted using a batch size of 2, with an initial learning rate of 0.002 and an Adam optimizer. The learning rate is adjusted over the course of training using a cosine annealing scheduler, which helps in fine-tuning the model parameters towards the latter stages of training. The models are trained for 100,000 optimization steps with segments of data of length \( L = 512 \) extracted from the superframes as described. Training is performed on NVIDIA A40 GPUs, and the complete training process takes approximately 23 hours for our proposed models. This setup ensures that the models are thoroughly optimized across the varied scenarios presented in the training dataset.
    
    \subsubsection{Evaluation Metrics}
    The performance of the models is evaluated using the \ac{MSE} score, a standard metric for quantifying the accuracy of models in regression tasks, which in this context measures the deviation of the model's outputs from the true signal values as detailed in \ref{sub:error_metrics}.
       
\section{Results}
\label{sec:results}
In this section, we present the results of our evaluations to demonstrate the performance and computational efficiency of the proposed architectures compared to the baseline models. Our analysis includes various aspects such as MSE score, computational complexity, inference speed, memory footprint, and parallelizability on GPUs.

We begin by examining the relationship between the \ac{MSE} score and the number of \ac{MACs}. This analysis allows us to evaluate the performance and computational resources required for each of the two proposed architectures as well as the two baseline models.
Following this, we present the \ac{MSE} score versus symbol rate of the model, which measures the inference speed and indicates how fast each architecture operates on both CPU and GPU. This is crucial for understanding the real-time applicability of the models.
Next, we provide a summary of the models, including their \ac{MSE} scores, \ac{MACs}, number of parameters, and size or memory footprint. We also discuss the impact of applying depthwise convolution and quantization techniques on these metrics.
Furthermore, we analyze the symbol rate of different models on a GPU for various batch sizes during inference. This analysis showcases the parallelizability of the architectures on GPUs, highlighting their efficiency in handling large-scale data processing.
Lastly, we explore the effects of applying pruning to the models to investigate if further compression and efficiency gains can be achieved through this method.

\subsection{Model Performance}

\subsubsection{Performance and Complexity}

\begin{figure}[ht]
\centering
\includegraphics[width=\columnwidth]{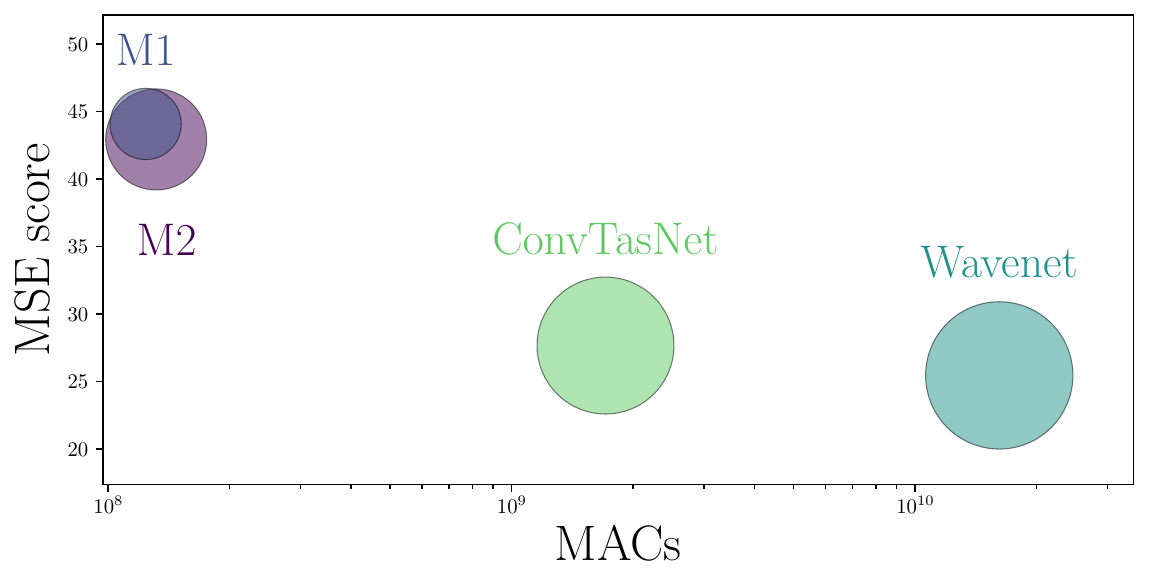}
\caption{Scatter plot of MSE score versus MACs for different models. The size of the circles is proportional to the number of parameters of each model.}
\label{fig:mse_vs_macs}
\end{figure}

To evaluate the performance and computational efficiency of our models, we present a scatter plot of the MSE score versus the number of \ac{MACs} in Fig. \ref{fig:mse_vs_macs}. This plot includes our two proposed architectures (M1 and M2) and two baseline models (ConvTasNet and WaveNet). The size of the circles in the plot is proportional to the number of parameters of each model, providing a visual representation of model complexity (computational as well as memory).

From Fig. \ref{fig:mse_vs_macs}, we observe the following. M1 and M2 achieve higher MSE scores compared to the baseline models, indicating superior performance in signal separation. Additionally, they require significantly fewer MACs, demonstrating lower computational complexity. The sizes of the circles for M1 and M2 indicate that both have fewer parameters than the baseline models, with M1 having slightly fewer parameters and performing slightly better in terms of MSE compared to M2.

ConvTasNet has a lower MSE score than M1 and M2, implying less effective signal separation. It also requires more MACs. The larger circle size indicates a higher number of parameters compared to M1 and M2. WaveNet has the lowest MSE score among all models, indicating the least effective performance in terms of signal separation. It demands the highest number of MACs and parameters, as shown by the large circle size. These findings underscore the trade-off between model performance and computational complexity. Although WaveNet requires substantial computational resources and has a high model complexity, it does not provide better signal separation compared to our proposed architectures. In contrast, M1 and M2 strike a balance by achieving superior signal separation with significantly lower computational demands, making them more suitable for deployment in resource-constrained environments.

Next, we evaluate the inference rate of these models to further understand their efficiency and practical applicability.

\subsubsection{Inference Rate}

The inference rate of the models is evaluated on both CPU and GPU, using the following hardware configurations:
\begin{itemize}
    \item \textbf{CPU:} Intel(R) Xeon(R) Silver 4310 CPU @ 2.10GHz with 48 cores.
     \item \textbf{GPU:} NVIDIA A40 with 46,068 MiB memory, CUDA Version 12.2.
\end{itemize}

The symbol rate is measured as the number of communication symbols that can be decoded when performing interference cancellation in one second. It is an important practical metric, as a low symbol rate can become a bottleneck for the end-to-end communication system, affecting overall system performance.

Before analyzing the symbol rate results, it is important to clarify a key point. Unlike \ac{MACs}, inference time is influenced not only by the model architecture and its theoretical computational complexity but also by the platform, which includes both the hardware (CPU or GPU) and the software and libraries that determine how layers (e.g., convolution) are implemented. This means that a model with fewer operations (in terms of \ac{MACs}) can still be slower depending on its implementation. In this work, we use PyTorch version 2.2.1+cu121 for both training and inference time measurements. The inference time measurements presented in this part of the study use a batch size of 1.
    \begin{figure}[ht]
        \centering
        \begin{minipage}{1.0\columnwidth}
            \centering
            \includegraphics[width=\columnwidth]{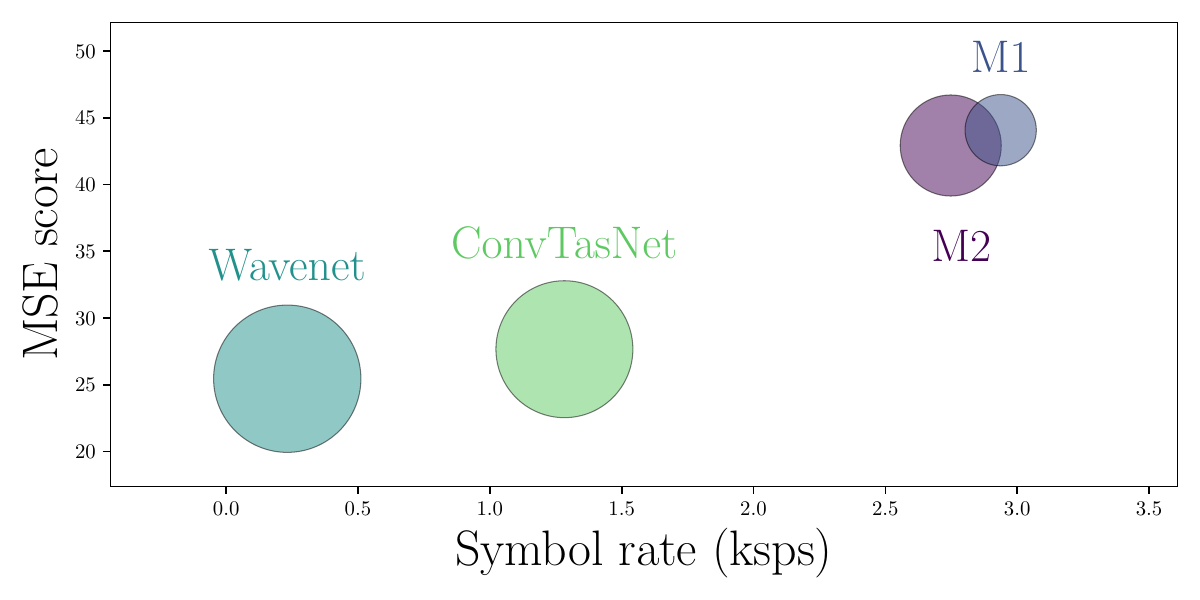}
            \caption{Performance comparison of different models on a CPU in terms of MSE score and symbol rate (ksps) with an inference batch size of one. The size of each bubble represents the model's number of parameters. M1 (U-Net with LSTM in the bottleneck) and M2 (fully convolutional U-Net) show higher symbol rates and better MSE scores compared to Wavenet and ConvTasNet.}
            \label{fig:MSE_MACs_cpu}
        \end{minipage}\\ 
    
        \begin{minipage}{1.0\columnwidth}
            \centering
            \includegraphics[width=\columnwidth]{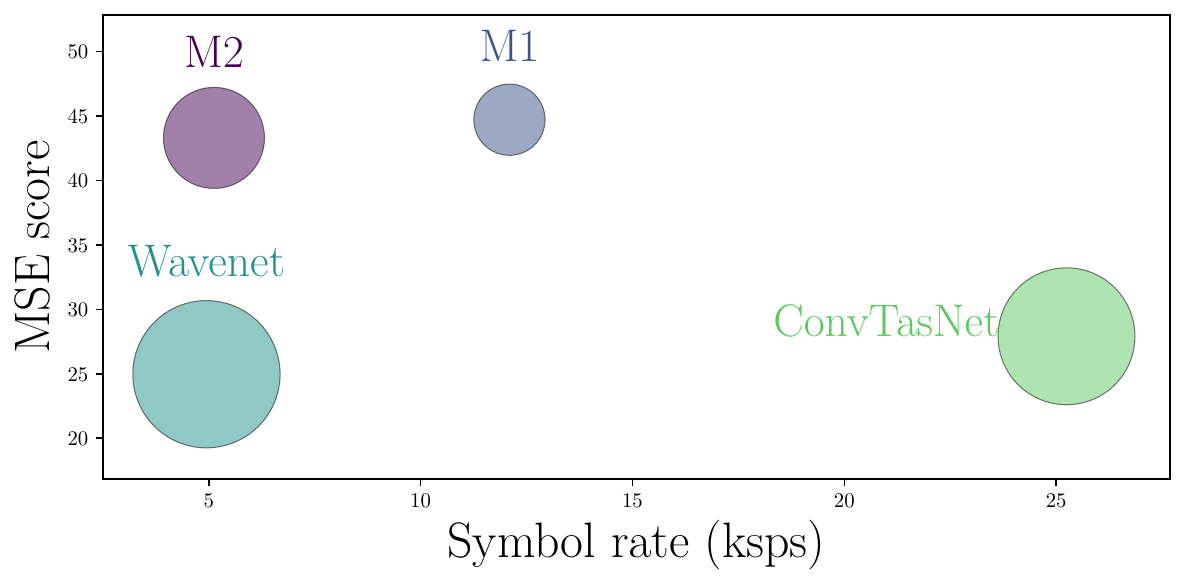}
            \caption{Performance comparison of different models on a GPU in terms of MSE score and symbol rate (ksps) with an inference batch size of one. The size of each bubble represents the model's number of parameters. On the GPU, ConvTasNet achieves a high symbol rate with a competitive MSE score, while M1 and M2 demonstrate high symbol rates and strong MSE scores. Wavenet shows moderate performance in both metrics.}
            \label{fig:MSE_MACs_cuda}
        \end{minipage}
    \end{figure}

From Fig. \ref{fig:MSE_MACs_cpu}, we observe the following:
On the CPU, M1 and M2 both achieve higher symbol rates compared to the baseline models, with M1 slightly outperforming M2. This indicates that our proposed architectures are more efficient in terms of inference speed on the CPU. As expected, having fewer MACs, both M1 and M2 demonstrate superior performance due to their efficient implementation.

However, on the GPU, ConvTasNet outperforms all other models in terms of symbol rate for inference batch size of one. This suggests that ConvTasNet benefits significantly from the parallel processing capabilities of the GPU (we will show it is not the case for larger batch sizes in section \ref{sec:parallelizability}). Additionally, M1 and M2 show a notable divergence in performance; M1 outperforms M2 in both symbol rate and MSE score. This highlights the superior scalability and efficiency of M1 when leveraging GPU resources compared to M2.
These observations underscore the importance of considering both computational complexity and platform-specific implementation when evaluating model performance. While our proposed models (M1 and M2) are highly efficient on the CPU, ConvTasNet demonstrates exceptional performance on the GPU, making it a viable option for high-throughput applications when a lower performance can be tolerated.


\subsection{Comparative Analysis}            
As shown in Table \ref{tab:model_metrics}, we provide a summary of the models' \ac{MSE} scores, \ac{MACs}, number of parameters, and memory footprint (SizeMb) with and without applying depthwise convolution (Dw) and quantization (Q).
\begin{table}[h!]
    \centering
    \begin{tabular}{lccccc}
        \toprule
        Model & MSE score & MACs  & Parameters & SizeMb \\
        \midrule
        M1 & \textbf{44.07} & 123,994,112  & 927,874 & 3.78 \\
        M1(Dw) & 43.75 & 51,252,224  & \textbf{514,570} & 2.13 \\
        M1(Q) & 33.10 & 123,994,112  & 927,874 & \textbf{1.90} \\
        M2 & 42.92 & 131,694,592  & 1,866,370 & 7.56 \\
        M2(Dw) & 43.19 & \textbf{51,231,232}  & 646,922 & 2.70 \\
        M2(Q) & 31.25 & 131,694,592  & 1,866,370 & 2.02 \\
        WaveNet & 25.45 & 16,175,333,376 & 3,964,674 & 15.92 \\
        ConvTasNet & 27.66 & 1,710,026,752 & 3,425,458 & 13.83 \\
        \bottomrule
    \end{tabular}
    \caption{Summary of Model Metrics}
    \label{tab:model_metrics}
\end{table}
The observed performance changes when applying depthwise (Dw) convolutions to U-Net architectures can be explained by the distinct roles of their layers. In the pure convolutional U-Net (M2), the reduction in parameters from Dw convolutions mitigates overfitting and promotes more efficient feature extraction, leading to a slight performance improvement from an \ac{MSE} score of 42.92 to 43.19. Conversely, in the U-Net with \ac{LSTM} (M1), the \ac{LSTM} layer depends on rich, high-dimensional feature maps to capture temporal dependencies effectively. Depthwise convolutions reduce the richness and informativeness of these feature maps, impairing the \ac{LSTM}'s ability to model temporal patterns, resulting in a slight performance degradation from an \ac{MSE} score of 44.07 to 43.75.

Quantization, which reduces precision to 8 bits, has a more severe impact on performance compared to using depthwise convolutions. The \ac{MSE} score drops significantly to 33.10 for M1 and to 31.25 for M2. However, quantization does not reduce the number of operations, and hence the \ac{MACs} remain unchanged.

The SizeMb column in Table \ref{tab:model_metrics} indicates the memory footprint required to store the model's weights. Depthwise convolutions reduce the \ac{MACs} to less than half in both M1 and M2, significantly decreasing the memory footprint to 2.13 MB for M1(Dw) and to 2.70 MB for M2(Dw). Quantization further reduces the model size to 1.90 MB for M1(Q) and to 2.02 MB for M2(Q).

These results highlight the trade-offs between model size, computational complexity, and performance. While our proposed architectures (M1 and M2) maintain a balance of high performance and low computational demands, applying depthwise convolutions and quantization can further optimize their memory footprint and computational efficiency. This makes them more suitable for deployment in resource-constrained environments while still achieving effective signal separation.

\subsection{Parallelizability on GPU}
\label{sec:parallelizability}
\begin{figure}[ht]
    \centering
    \includegraphics[width=\columnwidth]{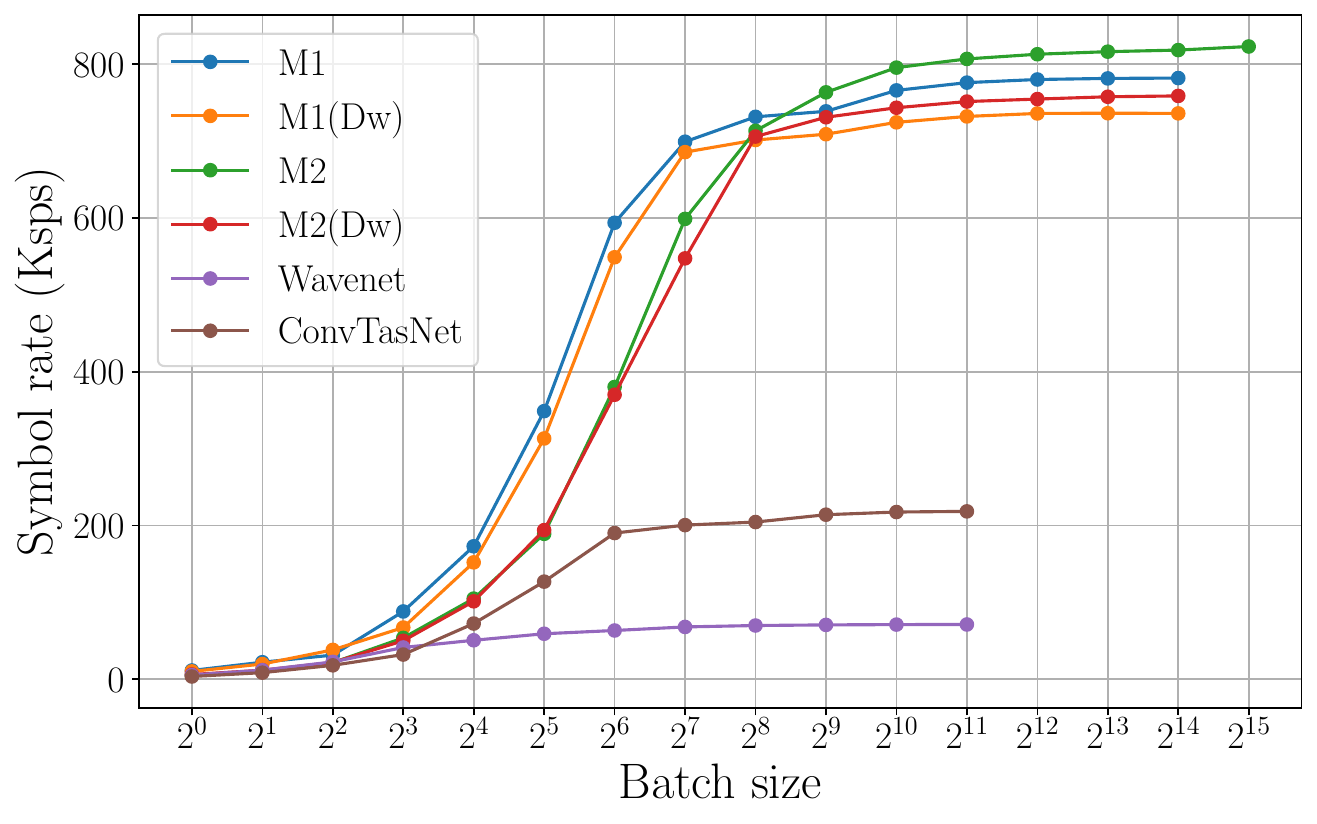}
    \caption{This graph demonstrates the increasing inference throughput when batch size is expanded, showcasing the scalability and efficiency gains of the model during batch processing.}
    \label{fig:SymRate_Batch}
    \end{figure}

Next, we analyze the symbol rate of different models on a GPU for various batch sizes during inference to showcase the parallelizability of the architectures. To evaluate the parallelizability of the models, we analyzed their performance on a GPU (NVIDIA A40 with 46,068 MiB memory, CUDA Version 12.2) across different batch sizes. The symbol rate (Ksps) as a function of batch size is depicted in Fig. \ref{fig:SymRate_Batch}. The purpose of this analysis is to determine how effectively each architecture can leverage the GPU's capability to process multiple batches of input simultaneously. It is important to note that we do not report quantization results for GPU since 8-bit precision computation is not typically supported on GPUs. Consequently, the (Q) models are omitted in this analysis.

From Fig. \ref{fig:SymRate_Batch}, several key observations can be made:
\begin{itemize}
    \item Initially, M1 and M1(Dw) achieve a higher symbol rate compared to M2. However, as the batch size increases beyond \(2^{10}\), the purely convolutional model M2 reaches the highest symbol rate at large batch sizes. This indicates that while models with \ac{LSTM} layers (M1 and M1(Dw)) perform well at smaller batch sizes, fully convolutional models (M2 and M2(Dw)) scale better with increasing batch sizes.
    \item WaveNet and ConvTasNet encounter \ac{OOM} errors at smaller batch sizes compared to M1, M1(Dw), M2, and M2(Dw), whereas M2 can handle the largest batch sizes before encountering \ac{OOM}  errors. This demonstrates the superior scalability of M2 in terms of batch processing on the GPU.
\end{itemize}

These results provide valuable insights for applications demanding high throughput or low latency. In such scenarios, the ability to process larger batch sizes effectively can significantly impact overall performance. As expected, models incorporating \ac{LSTM} layers (M1 and M1(Dw)) do not scale as well as fully convolutional models (M2 and M2(Dw)). Therefore, for applications where large batch processing and scalability are crucial, purely convolutional models like M2 are more advantageous.
In the following section, we explore the potential for further model compression through pruning and its impact on performance.

\subsection{Impact of Pruning}

To evaluate the potential for further model compression, we applied unstructured pruning to our models at various ratios and measured the resulting \ac{MSE} scores. Pruning is a technique used to reduce the number of parameters in a model, thereby decreasing its size. However, it can also degrade model performance if the pruned parameters are crucial for maintaining accuracy.

In this work, we apply unstructured pruning which removes individual weights from the convolutional and transposed convolutional layers based on their norm (pruning by removing filters with the lowest $\textit{l}_1$-norm), without considering their spatial positions. This type of pruning can lead to sparse weight matrices, but does not necessarily translate to faster inference since most hardware and software frameworks are optimized for dense computations.
\begin{table}[ht]
\centering
\caption{MSE Scores for Different Models and Pruning Ratios}
\begin{tabular}{lccccc}
\toprule
\multirow{2}{*}{Model} & \multicolumn{5}{c}{Pruning Ratio} \\
& 0.01 & 0.05 & 0.10 & 0.25 & 0.50 \\
\midrule
M1 & 31.67 & 30.32 & 25.69 & 15.94 & 6.51 \\
M1(Dw) & 30.59 & 26.67 & 20.49 & 10.21 & 2.14 \\
M2 & 29.62 & 28.85 & 24.34 & 16.19 & 7.42 \\
M2(Dw) & 28.05 & 25.50 & 20.30 & 10.79 & 2.43 \\
Wavenet & 12.51 & 12.54 & 12.34 & 10.50 & 6.14 \\
ConvTasNet & 14.12 & 13.67 & 12.17 & 6.88 & 1.96 \\
\bottomrule
\end{tabular}
\label{tab:pruning_results}
\end{table}

Table \ref{tab:pruning_results} presents the \ac{MSE} scores for different models and pruning ratios. From these results, we observe a general trend of performance degradation as the pruning ratio increases. This is expected, as higher pruning ratios remove more parameters, which can negatively impact the model's ability to accurately separate signals.

For M1, the \ac{MSE} score decreases from 31.67 to 6.51 as the pruning ratio increases from 0.01 to 0.50. Similarly, M1(Dw) shows a significant drop in performance, with the \ac{MSE} score decreasing from 30.59 to 2.14. This suggests that M1 and M1(Dw) are highly sensitive to pruning, likely due to their reliance on rich feature representations to capture temporal dependencies.

M2 and M2(Dw) exhibit a similar sensitivity to pruning. The \ac{MSE} score for M2 drops from 29.62 to 7.42, while M2(Dw) decreases from 28.05 to 2.43. The pure convolutional architecture (M2) and its depthwise variant  also rely on a large number of parameters to perform effective feature extraction, which explains their performance degradation with increased pruning.

Additionally, the reference models ConvTasNet and Wavenet have been included to offer a comparative perspective. ConvTasNet starts with an \ac{MSE} of 14.12 at the lowest pruning ratio of 0.01, dropping to 1.96 at 0.50, showing its robustness compared to other models. Wavenet begins with an \ac{MSE} of 12.51 at 0.01 and decreases to 6.14 at 0.50, indicating a relatively stable performance across varying pruning levels.

The pronounced decline in \ac{MSE} scores across all models underscores the substantial impact of pruning on performance. This is particularly notable when compared with the optimal scores previously reported, highlighting a significant degradation. The sensitivity observed can be attributed to the extensive hyperparameter tuning tailored to maximize model efficiency under fully-parameterized conditions, inadvertently rendering the models less tolerant to the reduction of parameters.

In practical applications, this sensitivity implies that while some degree of model compression is feasible, pruning in non-overparameterized models can lead to considerable performance losses. Consequently, careful consideration must be given to the trade-off between model size reduction and performance degradation when employing pruning techniques for model compression.

This concludes our evaluation of model performance, inference rate, parallelizability, and the impact of pruning. These comprehensive analyses provide valuable insights into the strengths and limitations of the proposed architectures and their suitability for real-world deployment.
\section{Discussion}
\label{sec:discussion}
\subsection{Analysis of Results}
The results presented in this study provide an evaluation of the proposed architectures (M1 and M2) compared to the baseline models (ConvTasNet and WaveNet). Our analysis indicates that M1 and M2 achieve higher \ac{MSE} scores with significantly fewer \ac{MACs}, demonstrating their efficiency in terms of computational complexity. Specifically, M1 and M2 showed strong performance in signal separation while maintaining a lower computational footprint, making them suitable for deployment in resource-constrained environments.

When evaluating the inference rate on the GPU, we observed that M1 initially achieves higher symbol rates at smaller batch sizes. However, as batch size increases, M2 outperforms M1, indicating better scalability and parallelizability. ConvTasNet, despite its higher computational demands, demonstrates exceptional performance in a single batch setup on GPU. These results highlight the importance of considering both computational complexity and implementation efficiency when selecting models for real-time applications.

The application of depthwise convolutions and quantization further reduced the memory footprint and computational demands of M1 and M2. However, this came at the cost of some performance degradation. Depthwise convolutions, in particular, had a varied impact, slightly improving the performance of M2 while degrading M1. Quantization led to a more significant drop in performance, underscoring the trade-offs involved in model optimization techniques.

The pruning results revealed that our models are sensitive to parameter reduction, likely due to extensive hyperparameter tuning during their design. The significant performance degradation observed with higher pruning ratios suggests that pruning may not be the most effective strategy for optimizing these models for faster inference.

\subsection{Deployment Considerations}

Given the focus on edge devices, we have expanded our discussion to include deployment-related considerations such as power consumption, latency, and hardware compatibility. Power consumption is linked to the number of MACs, as established in studies like \cite{8335698}, and our proposed models significantly reduce MACs through the use of depthwise separable convolutions and quantization, achieving up to a 60\% reduction in computational demands. Latency has been evaluated in terms of inference speed, with the fully convolutional model (M2) demonstrating superior scalability and throughput, achieving up to $800,000$ symbols per second on GPUs. Additionally, hardware compatibility is addressed through quantization-aware training (QAT), which reduces model size (e.g., M1(Q): $1.90$ MB) and ensures efficient deployment on low-precision devices without significant performance degradation. These considerations underscore the practical viability of our models for real-time applications on resource-constrained platforms.

\subsection{Implications and Future Work}

The findings from this study have several practical implications. The superior performance and lower computational complexity of M1 and M2 make them attractive for applications where computational resources are limited. Their ability to maintain high \ac{MSE} scores with fewer \ac{MACs} highlights their potential for efficient real-time signal processing in communication systems.

The varying scalability of the models on the GPU suggests that model selection should be tailored to the specific requirements of the deployment environment. For instance, applications demanding high throughput and low latency may benefit more from M2, given its superior performance at larger batch sizes.

Future work will focus on further optimizing these models for deployment. This includes exploring structured pruning techniques that may offer a better trade-off between performance and computational efficiency. Additionally, investigating more advanced quantization methods that minimize performance loss while reducing model size will be a key area of research. Enhancing the implementation of sparse operations to leverage hardware capabilities more effectively can also provide significant performance improvements.

Moreover, extending the evaluation to include other relevant metrics, such as \ac{BER} analysis, will provide a more comprehensive understanding of the models' performance in practical communication scenarios. Finally, adapting these models to different hardware platforms and exploring their potential in other signal processing applications will be crucial for broadening their applicability.

\section{Conclusion}
\label{sec:conclusion}
In this paper, we have presented an in-depth evaluation of two proposed U-Net architectures (M1 and M2) for signal separation, comparing them with baseline models (ConvTasNet and WaveNet). Our results demonstrate that M1 and M2 achieve superior performance in terms of \ac{MSE} scores while maintaining significantly lower computational complexity, making them suitable for deployment in resource-constrained environments.

Specifically, M1, which incorporates depthwise separable convolutions and an LSTM, reduces MACs by 58.66\% with only a 0.72\% degradation in \ac{MSE} score. M2, a fully convolutional model, when implemented by depthwise separable convolutions achieves a 0.63\% improvement in \ac{MSE} score while using 61.10\% fewer MACs compared to original M2  models. These substantial reductions in computational complexity highlight the efficiency of the proposed models.

Our analysis of inference rates on both CPU and GPU indicates that M2 scales particularly well with larger batch sizes, achieving symbol rates of up to 800,000 symbols per second. This scalability makes M2 an excellent choice for applications demanding high throughput and low latency. The optimization techniques, including quantization and depthwise separable convolutions, effectively reduce memory footprint and computational demands, though some performance trade-offs are observed. Pruning, especially unstructured pruning, demonstrated the models' sensitivity to parameter reduction, suggesting the need for more sophisticated optimization strategies.

These findings provide valuable insights into selecting and optimizing models for real-time signal processing applications. Future research will explore structured pruning techniques, advanced quantization methods, and leveraging hardware capabilities for sparse operations. Additionally, extending the evaluation to include BER analysis and adapting the models to various hardware platforms will further enhance their practical applicability.

In conclusion, our proposed architectures achieve a promising balance between performance and efficiency, positioning them as strong candidates for advanced signal separation tasks in communication systems and beyond.

\section{Acknowledgments}
This work was supported in part by the European Union through the Horizon Europe Marie Sklodowska-Curie Staff Exchange programme “Electric Vehicles Point Location Optimisation via Vehicular Communications (EVOLVE),” under Grant 101086218, and in part by the U.S. National Science Foundation under Grant ECCS-2335876.

\bibliographystyle{IEEEtran}
\bibliography{main}
\end{document}